\documentclass[english,american,aps,prb,twocolumn,floatfix,showpacs]{revtex4}
\usepackage[latin1]{inputenc}
\usepackage{graphicx}
\usepackage{amssymb}

\makeatletter

\usepackage{color}
\usepackage{graphicx}
\usepackage{epsfig}
\usepackage{amsbsy}

\usepackage{babel}
\makeatother
\begin{document}

\newcommand{\commute}[2]{\left[#1,#2\right]}

\newcommand{\bra}[1]{\left\langle #1\right|}

\newcommand{\ket}[1]{\left|#1\right\rangle }

\newcommand{\anticommute}[2]{\left\{  #1,#2\right\}  }
\foreignlanguage{english}{}

\title{Singlet-triplet decoherence due to nuclear spins in a double quantum
dot}

\author{W. A. Coish}

\author{Daniel Loss}

\affiliation{Department of Physics and Astronomy, University of Basel, Klingelbergstrasse
82, CH-4056 Basel, Switzerland}

\begin{abstract}
We have evaluated hyperfine-induced electron spin dynamics for two
electrons confined to a double quantum dot. Our quantum solution accounts
for decay of a singlet-triplet correlator even in the presence of
a fully static nuclear spin system, with no ensemble averaging over
initial conditions. In contrast to an earlier semiclassical calculation,
which neglects the exchange interaction, we find that the singlet-triplet
correlator shows a long-time saturation value that differs from $1/2$,
even in the presence of a strong magnetic field. Furthermore, we find
that the form of the long-time decay undergoes a transition from a
rapid Gaussian to a slow power law ($\sim1/t^{3/2}$) when the exchange
interaction becomes nonzero and the singlet-triplet correlator acquires
a phase shift given by a universal (parameter independent) value of
$3\pi/4$ at long times. The oscillation frequency and time-dependent
phase shift of the singlet-triplet correlator can be used to perform
a precision measurement of the exchange interaction and Overhauser
field fluctuations in an experimentally accessible system. We also
address the effect of orbital dephasing on singlet-triplet decoherence,
and find that there is an optimal operating point where orbital dephasing
becomes negligible. 
\end{abstract}
\maketitle

\section{Introduction}

Decoherence due to the coupling of a qubit to its environment is widely
regarded as the major obstacle to quantum computing and quantum information
processing in solid-state systems. Electron spins confined in semiconductor
quantum dots\cite{loss:1998a} couple to their environments primarily
through the spin-orbit interaction and hyperfine interaction with
nuclear spins in the surrounding lattice.\cite{burkard:1999a,cerletti:2004a}
To reach the next step in coherent electron spin state manipulation,
the strongest decoherence effects in this system must be understood
and reduced, if possible. 

The effects of spin-orbit interaction are reduced in confined quantum
dots at low temperatures.\cite{khaetskii:2000a} Indeed, recent experiments
give longitudinal relaxation times $T_{1}$ for quantum-dot-confined
electrons that reach $T_{1}\approx20\,\mathrm{ms}$\cite{kroutvar:2004a}
in self-assembled dots and $T_{1}\approx0.85\,\mathrm{ms}$ in gated
dots\cite{elzerman:2004a}, in agreement with theory.\cite{golovach:2004a}
These times suggest that the spin-orbit interaction is a relatively
weak source of decoherence in these structures since theory predicts
that the transverse spin decay time $T_{2}$ due to spin-orbit interaction
alone (neglecting other sources of decoherence) would be given by
$T_{2}=2T_{1}$.\cite{golovach:2004a} Other strategies for reducing
the effects of spin-orbit interaction may include using hole (instead
of electron) spin, where a recent study has found that $T_{2}=2T_{1}$
also applies, and the hole spin relaxation time can be made even longer
than that for the electron spin.\cite{bulaev:2004a} 

Unlike the spin-orbit interaction, the hyperfine interaction of a
single electron spin with a random nuclear spin environment can lead
to pure dephasing, giving a transverse spin decay time on the order
of $5\,\mathrm{ns}$,\cite{khaetskii:2002a,merkulov:2002a,coish:2004a}
six orders of magnitude shorter than the measured longitudinal decay
times $T_{1}$. To minimize errors during qubit gating operations
in these proposed devices, this decay must be fully understood. The
hyperfine interaction in a single quantum dot is described by a Hamiltonian
$H=\mathbf{h}\cdot\mathbf{S}$, where $\mathbf{S}$ is the electron
spin operator and $\mathbf{h}$ is a collective quantum nuclear spin
operator, which we will refer to as the {}``Overhauser operator''.
A common assumption in the literature is to replace the Overhauser
operator by a classical effective magnetic field $\mathbf{h}\to\mathbf{B}_{N}$.\cite{schulten:1978a,merkulov:2002a,khaetskii:2002a,erlingsson:2003a,erlingsson:2004a,yuzbashyan:2004a,bracker:2005a,braun:2005a,gurudevdutt:2005a,taylor:2005a,petta:2005a,koppens:2005a,greilich:2005a}
Since a classical magnetic field only induces precession (not decoherence),
the classical-field picture necessitates an ensemble of nuclear spin
configurations to induce decay of the electron spin expectation value.\cite{khaetskii:2002a,merkulov:2002a}
For experiments performed on a large bulk sample of electron spins,
or experiments performed over timescales that are longer than the
typical timescale for variation of $\mathbf{B}_{N}$, the source of
the ensemble averaging is clear. However, one conclusion of this model
is that single-electron-spin experiments performed over a timescale
shorter than the nuclear spin correlation time should show no decay.
This conclusion is contradicted by numerical \cite{schliemann:2002a,shenvi:2005a}
and analytical \cite{zurek:2003a,coish:2004a} results, which show
that the quantum nature of the Overhauser operator can lead to rapid
decay of a single electron spin, even for a fully static nuclear spin
system. This rapid decay is, however, reversible with a standard Hahn
spin-echo sequence in an applied magnetic field and the timescale
of the decay can be increased by squeezing the nuclear spin state.\cite{coish:2004a} 

Another potential solution to the hyperfine decoherence problem is
to polarize the nuclear spins. Polarizing the nuclear spin system
in zero applied magnetic field reduces the longitudinal spin-flip
probability by the factor $1/p^{2}N$, where $p$ is the nuclear spin
polarization and $N$ is the number of nuclear spins within the quantum
dot.\cite{burkard:1999a,coish:2004a} The effect on the transverse
components of electron spin is different. Unless the nuclear spin
state is squeezed or a spin-echo sequence is performed, the transverse
components of electron spin will decay to zero in a time $t_{c}\approx5\,\mathrm{ns}$
in a typical GaAs quantum dot. Polarizing the nuclear spin system
increases $t_{c}$ by reducing the phase-space available for fluctuations
in the Overhauser operator, resulting in $t_{c}\approx5\,\mathrm{ns}/\sqrt{1-p^{2}}$.\cite{coish:2004a}
Recent experiments show that the nuclear spin system can be polarized
by as much as 60\%.\cite{bracker:2005a} However, to achieve an order-of-magnitude
increase in $t_{c}$, the polarization degree would have to be on
the order of 99\%,\cite{cerletti:2004a} for which more ambitious
polarization schemes have been proposed.\cite{imamoglu:2003a} 

If electron spins in quantum dots are to be used as quantum information
processors, the two-electron states of double quantum dots must also
be coherent during rapid two-qubit switching times.\endnote{For exchange gates with spin-1/2 qubits\cite{loss:1998a}, the relevant requirement is that the qubit switching time $t_S$ should be much smaller than the singlet-triplet decoherence time.\cite{burkard:1999a}}
Measurements of singlet-triplet relaxation times $t_{ST}$ in vertical
double dots $(t_{ST}\approx200\,\mu\mathrm{s})$,\cite{fujisawa:2002a}
gated lateral double dots $(t_{ST}\approx70\,\mu\mathrm{s})$,\cite{petta:2004a}
and single dots $(t_{ST}\approx2.58\,\mathrm{ms})$\cite{hanson:2005a}
suggest that these states may be very long-lived. Recent experiments
have now probed the decoherence time of such states, which is believed
to be limited by the hyperfine interaction with surrounding nuclear
spins.\cite{petta:2005a} The dramatic effect of the hyperfine interaction
on two-electron states in a double quantum dot has previously been
illustrated in experiments that show slow time-dependent current oscillations
in transport current through a double dot in the spin blockade regime.\cite{ono:2004a} 

It may be possible to circumvent some of the complications associated
with single-spin decoherence by considering an encoded qubit, composed
of the two-dimensional subspace of states with total $z$-projection
of spin equal to zero for two electrons in a double quantum dot.\cite{taylor:2005a}
One potential advantage of such a setup is that it may be possible
to reduce the strength of hyperfine coupling to the encoded state
space for a symmetric double-dot (see Appendix \ref{sec:Overhauserfieldestimates}).
A potential disadvantage of this scheme is that coupling to the orbital
(charge) degree of freedom can then lead to additional decoherence,
but we find that orbital dephasing can be made negligible under appropriate
conditions (see Sec. \ref{sec:Orbital-dephasing}). To achieve control
of the singlet-triplet subspace, however, the decoherence process
for the two-electron system should be understood in detail. 

In this paper we give a fully quantum mechanical solution for the
spin dynamics of a two-electron system coupled to a nuclear-spin environment
via the hyperfine interaction in a double quantum dot. Although we
focus our attention here on quantum dots, decoherence due to a spin
bath is also an important problem for, e.g., proposals to use molecular
magnets for quantum information processing.\cite{cerletti:2004a,meier:2003a,meier:2003b,troiani:2005a}
In fact, the problem of a pair of electrons interacting with a bath
of nuclear spins via the contact hyperfine interaction has been addressed
long ago to describe spin-dependent reaction rates in radicals.\cite{schulten:1978a,werner:1977a}
A semiclassical theory has been developed,\cite{schulten:1978a} in
which electron spins in radicals experience a randomly oriented effective
classical magnetic field due to the contact hyperfine interaction
between electron and nuclear spins. In this semiclassical theory,
random hopping events of the electrons were envisioned to induce a
randomly fluctuating local magnetic field at the site of the electron
spin, resulting in decay of a singlet-triplet correlator. Here, we
solve a different problem. Ensemble averaging over nuclear spin configurations
is natural for a large sample of $\sim10^{23}$ radicals. In contrast,
we consider the coherent dynamics of two-electron spin states within
a single double quantum dot. More importantly, the Heisenberg exchange
interaction, which was found to be negligible in Ref. \onlinecite{schulten:1978a},
can be any value (large or small) in our system of interest. We find
that a nonzero exchange interaction can lead to a drastic change in
the form and timescale of decoherence. Moreover, this paper is of
direct relevance to very recent experiments \cite{johnson:2005a,petta:2005a,koppens:2005a}
related to such double-dot systems. 

The rest of this paper is organized as follows. In Sec. \ref{sec:SzZero}
we solve the problem for electron spin dynamics in the subspace of
total spin $z$-component $S^{z}=0$ with an exact solution for the
projected effective Hamiltonian. In Sec. \ref{sec:STplus} we show
that a perturbative solution is possible for electron spin dynamics
in the subspace of singlet and $S^{z}=+1$ triplet states. Sec. \ref{sec:Orbital-dephasing}
contains a discussion of the contributions to singlet-triplet decoherence
from orbital dephasing. In Sec. \ref{sec:Conclusions} we review our
most important results. Technical details are given in Appendixes
\ref{sec:Overhauserfieldestimates} to \ref{sec:Asymptotics}.

\section{\label{sec:SzZero}Dynamics in the $S^{z}=0$ subspace}

We consider two electrons confined to a double quantum dot, of the
type considered, for example, in Refs. \onlinecite{johnson:2005a,petta:2005a,koppens:2005a}.
Each electron spin experiences a Zeeman splitting $\epsilon_{z}=g\mu_{B}B$
due to an applied magnetic field $\mathbf{B}=(0,0,B)$, $B>0$, defining
the spin quantization axis $z$, which can be along or perpendicular
to the quantum dot axis. In addition, each electron interacts with
an independent quantum nuclear field $\mathbf{h}_{l},\, l=1,2$, due
to the contact hyperfine interaction with surrounding nuclear spins.
The nuclear field experienced by an electron in orbital state $l$
is $\mathbf{h}_{l}=\sum_{k}A_{k}^{l}\mathbf{I}_{k}$, where $\mathbf{I}_{k}$
is the nuclear spin operator for a nucleus of total spin $I$ at lattice
site $k$, and the hyperfine coupling constants are given by $A_{k}^{l}=vA\left|\psi_{0}^{l}(\mathbf{r}_{k})\right|^{2}$,
with $v$ the volume of a unit cell containing one nuclear spin, $A$
characterizes the hyperfine coupling strength, and $\psi_{0}^{l}(\mathbf{r}_{k})$
is the single-particle envelope wavefunction for orbital state $l$,
evaluated at site $k$. This problem simplifies considerably in a
moderately large magnetic field ($B\gg\mathrm{max\{\left\langle \delta\mathbf{h}\right\rangle _{\mathrm{rms}}/g\mu_{B},\left\langle \mathbf{h}\right\rangle _{\mathrm{rms}}/g\mu_{B}\}}$,
where $\left\langle \mathcal{O}\right\rangle _{\mathrm{rms}}=\bra{\psi_{I}}\mathcal{O}^{2}\ket{\psi_{I}}^{1/2}$
is the root-mean-square expectation value of the operator $\mathcal{O}$
with respect to the nuclear spin state $\ket{\psi_{I}}$, $\delta\mathbf{h}=\frac{1}{2}\left(\mathbf{h}_{1}-\mathbf{h}_{2}\right)$,
and $\mathbf{h}=\frac{1}{2}\left(\mathbf{h}_{1}+\mathbf{h}_{2}\right)$).
In a typical unpolarized GaAs quantum dot, this condition is $B\gg IA/\sqrt{N}g\mu_{B}\approx10\,\mathrm{mT}$
(see Appendix \ref{sec:Overhauserfieldestimates}). For this estimate,
we have used $IA/g\mu_{\mathrm{B}}\approx5\,\mathrm{T}$, based on
a sum over all three nuclear spin isotopes (all three hyperfine coupling
constants) present in GaAs\cite{paget:1977a} and $N\approx10^{5}$
nuclei within each quantum dot. In this section, we also require $B\gg J/g\mu_{B}$,
where $J$ is the Heisenberg exchange coupling between the two electron
spins. For definiteness we take $J>0$, but all results are valid
for either sign of $J$, with $J$ replaced by its absolute value.
In the above limits, the electron Zeeman energy dominates all other
energy scales and the relevant spin Hamiltonian becomes block-diagonal,
with blocks labeled by the total spin projection along the magnetic
field $S^{z}$ (see Appendix \ref{sec:Effective-Hamiltonians}). In
the subspace of $S^{z}=0$ we write the projected two-electron spin
Hamiltonian in the subspace of singlet and $S^{z}=0$ triplet states
$(\ket{S},\ket{T_{0}})$ to zeroth order in the inverse Zeeman splitting
$1/\epsilon_{z}$ as $H_{0}=\frac{J}{2}\mathbf{S}\cdot\mathbf{S}+\delta h^{z}\delta S^{z}$,
where $\mathbf{S}=\mathbf{S}_{1}+\mathbf{S}_{2}$ is the total spin
operator in the double dot and $\delta\mathbf{S}=\mathbf{S}_{1}-\mathbf{S}_{2}$
is the spin difference operator. In terms of the vector of Pauli matrices
$\pmb{\tau}=(\tau^{x},\tau^{y},\tau^{z})$:$\,\,\,\ket{S}\to\ket{\tau^{z}=-1},\,\ket{T_{0}}\to\ket{\tau^{z}=+1}$
$H_{0}$ can be rewritten as:\begin{equation}
H_{0}=\frac{J}{2}\left(1+\tau^{z}\right)+\delta h^{z}\tau^{x}.\end{equation}
 Diagonalizing this two-dimensional Hamiltonian gives eigenvalues
and eigenvectors

\begin{eqnarray}
E_{n}^{\pm} & = & \frac{J}{2}\pm\frac{1}{2}\sqrt{J^{2}+4\left(\delta h_{n}^{z}\right)^{2}},\label{eq:LargeBEigenvals}\\
\ket{E_{n}^{\pm}} & = & \frac{\delta h_{n}^{z}\ket{S}+E_{n}^{\pm}\ket{T_{0}}}{\sqrt{\left(E_{n}^{\pm}\right)^{2}+\left(\delta h_{n}^{z}\right)^{2}}}\otimes\ket{n},\end{eqnarray}
 where $\ket{n}$ is an eigenstate of the operator $\delta h^{z}$
with eigenvalue $\delta h_{n}^{z}$. Since the eigenstates $\ket{E_{n}^{\pm}}$
are simultaneous eigenstates of the operator $\delta h^{z}$, we note
that there will be no dynamics induced in the nuclear system under
the Hamiltonian $H_{0}$. In other words, the nuclear system remains
static under the influence of $H_{0}$ alone, and there is consequently
no back action on the electron spin due to nuclear dynamics. 

We fix the electron system in the singlet state $\ket{S}$ at time
$t=0$:\begin{equation}
\ket{\psi(t=0)}=\ket{S}\otimes\ket{\psi_{I}};\,\,\,\,\,\ket{\psi_{I}}=\sum_{n}a_{n}\ket{n},\end{equation}
 where $a_{n}$ is an arbitrary set of (normalized) coefficients $(\sum_{n}\left|a_{n}\right|^{2}=1)$.
The initial nuclear spin state $\ket{\psi_{I}}$ is, in general, not
an eigenstate $\ket{n}$. The probability to find the electron spins
in the state $\ket{T_{0}}$ at $t>0$ is then given by the correlation
function (setting $\hbar=1$):\begin{equation}
C_{T_{0}}(t)=\sum_{n}\rho_{I}(n)\left|\bra{n}\otimes\bra{T_{0}}e^{-iH_{0}t}\ket{S}\otimes\ket{n}\right|^{2},\end{equation}
 where $\rho_{I}(n)=|a_{n}|^{2}$ gives the diagonal matrix elements
of the nuclear-spin density operator, which describes a pure (not
mixed) state of the nuclear system: $\rho_{I}=\ket{\psi_{I}}\bra{\psi_{I}}=\sum_{n}\rho_{I}(n)\ket{n}\bra{n}+\sum_{n\ne n^{\prime}}a_{n}^{*}a_{n^{\prime}}\ket{n^{\prime}}\bra{n}$.
$C_{T_{0}}(t)$ is the sum of a time-independent piece $\overline{C_{n}}$
and an interference term $C_{T_{0}}^{\mathrm{int}}(t)$:\begin{eqnarray}
C_{T_{0}}(t) & = & \overline{C_{n}}+C_{T_{0}}^{\mathrm{int}}(t),\\
C_{n} & = & \frac{2\left(\delta h_{n}^{z}\right)^{2}}{J^{2}+4\left(\delta h_{n}^{z}\right)^{2}},\label{eq:CnCoefficients}\\
C_{T_{0}}^{\mathrm{int}}(t) & = & -\overline{C_{n}\cos\left(\left[E_{n}^{+}-E_{n}^{-}\right]t\right)}.\label{eq:CintDiscrete}\end{eqnarray}
 Here, the overbar is defined by $\overline{f(n)}=\sum_{n}\rho_{I}(n)f(n)$.
Note that $C_{n}$ depends only on the exchange and Overhauser field
inhomogeneity $\delta h_{n}^{z}$ through the ratio $\delta h_{n}^{z}/J$. 

For a large number of nuclear spins $N\gg1$ in a superposition of
$\delta h^{z}$-eigenstates $\ket{n}$, we assume that $\rho_{I}(n)$
describes a continuous Gaussian distribution of $\delta h_{n}^{z}$
values, with mean $\overline{\delta h_{n}^{z}}=0$ (for the case $\overline{\delta h_{n}^{z}}\ne0$,
see Sec. \ref{sub:Deltahnonzero}) and variance $\sigma_{0}^{2}=\overline{\left(\delta h_{n}^{z}-\overline{\delta h_{n}^{z}}\right)^{2}}=\overline{\left(\delta h_{n}^{z}\right)^{2}}$
(i.e. $\sigma_{0}=\left\langle \delta h^{z}\right\rangle _{\mathrm{rms}}$).
The approach to a Gaussian distribution in the limit of large $N$
for a sufficiently randomized nuclear system is guaranteed by the
central limit theorem.\cite{coish:2004a} The assumption of a continuous
distribution of $\delta h_{n}^{z}$ precludes any possibility of recurrence
in the correlator we calculate.\endnote{We recall that a superposition $f(t)$ of oscillating functions with different periods leads to quasiperiodic behavior, i.e., after the so-called Poincar{\'e} recurrence time $t_p$, the function $f(t)$ will return back arbitrarily close to its initial value (see, e.g., Ref. \onlinecite{fick:1990a}).}
A lower-bound for the Poincar{\'e} recurrence time in this system
is given by the inverse mean level spacing for the fully-polarized
problem\cite{khaetskii:2002a}: $t_{\mathrm{p}}\gtrsim N^{2}/A$.
In a GaAs double quantum dot containing $N\simeq10^{5}$ nuclear spins,
this estimate gives $t_{\mathrm{rec}}\gtrsim0.1\,\mathrm{s}$. Moreover,
by performing the continuum limit, we restrict ourselves to the free-induction
signal (without spin-echo). In fact, we remark that \emph{all} decay
in the correlator given by (\ref{eq:CintDiscrete}) can be recovered
with a suitable $\pi$-pulse, defined by the unitary operation $U_{\pi}\ket{E_{n}^{\pm}}=\ket{E_{n}^{\mp}}$.
This statement follows directly from the sequence\begin{equation}
e^{-iJt}\ket{E_{n}^{\pm}}=U_{\pi}e^{-iH_{0}t}U_{\pi}e^{-iH_{0}t}\ket{E_{n}^{\pm}}.\end{equation}
 Thus, under the above sequence of echoes and free induction, all
eigenstates are recovered up to a common phase factor. Only higher-order
corrections to the effective Hamiltonian $H_{0}$ may induce completely
irreversible decay. This irreversible decay could be due, for example,
to the variation in hyperfine coupling constants, leading to decay
on a timescale $t\sim N/A$, as in the case of a single electron spin
in Refs. \onlinecite{khaetskii:2002a,coish:2004a}. Another source
of decay is orbital dephasing (see Sec. \ref{sec:Orbital-dephasing}). 

We perform the continuum limit for the average of an arbitrary function
$f(n)$ according to the prescription\begin{eqnarray}
\sum_{n}\rho_{I}(n)f(n) & \to & \int dxP_{\sigma;\overline{x}}(x)f(n(x)),\\
P_{\sigma;\overline{x}}(x) & = & \frac{1}{\sqrt{2\pi}\sigma}\exp\left(-\frac{\left(x-\overline{x}\right)^{2}}{2\sigma^{2}}\right),\end{eqnarray}
 with $\overline{x}=0$, $\sigma^{2}=\overline{x^{2}}$, and here
we take $x=\delta h_{n}^{z},\,\sigma=\sigma_{0}$. Using \begin{equation}
C_{n}=C(\delta h_{n}^{z})=C(x)=\frac{2x^{2}}{J^{2}+4x^{2}},\label{eq:Cofx}\end{equation}
 we evaluate $C_{T_{0}}^{\mathrm{int}}(t)=\mathrm{Re}\left[\tilde{C}_{T_{0}}^{\mathrm{int}}(t)\right]$,
where the complex interference term is given by the integral\begin{eqnarray}
\tilde{C}_{T_{0}}^{\mathrm{int}}(t) & = & -\int_{-\infty}^{\infty}dxC(x)P_{\sigma_{0};0}(x)e^{it\sqrt{J^{2}+4x^{2}}}.\label{eq:ComplexCT0}\end{eqnarray}

In general, the interference term given by Eq. (\ref{eq:ComplexCT0})
will decay to zero after the singlet-triplet decoherence time. We
note that the interference term decays \emph{even} for a \emph{purely
static} nuclear spin configuration with no ensemble averaging performed
over initial conditions, as is the case for an isolated electron spin.\cite{schliemann:2002a,zurek:2003a,coish:2004a}
The total $z$-component of the nuclear spins will be essentially
static in any experiment performed over a timescale less than the
nuclear spin diffusion time (the diffusion time is several seconds
for nuclei surrounding donors in GaAs\cite{paget:1982a}). We stress
that the relevant timescale in the present case is the spin diffusion
time, and not the dipolar correlation time, since nonsecular corrections
to the dipole-dipole interaction are strongly suppressed by the nuclear
Zeeman energy in an applied magnetic field of a few Gauss\cite{slichter:1980a}
(as assumed here). Without preparation of the initial nuclear state
or implementation of a spin-echo technique, this decoherence process
therefore cannot be eliminated with fast measurement, and in general
cannot be modeled by a classical nuclear field moving due to slow
internal dynamics; a classical nuclear field that does not move cannot
induce decay. 

\begin{figure}
\includegraphics[%
  clip,
  scale=0.65]{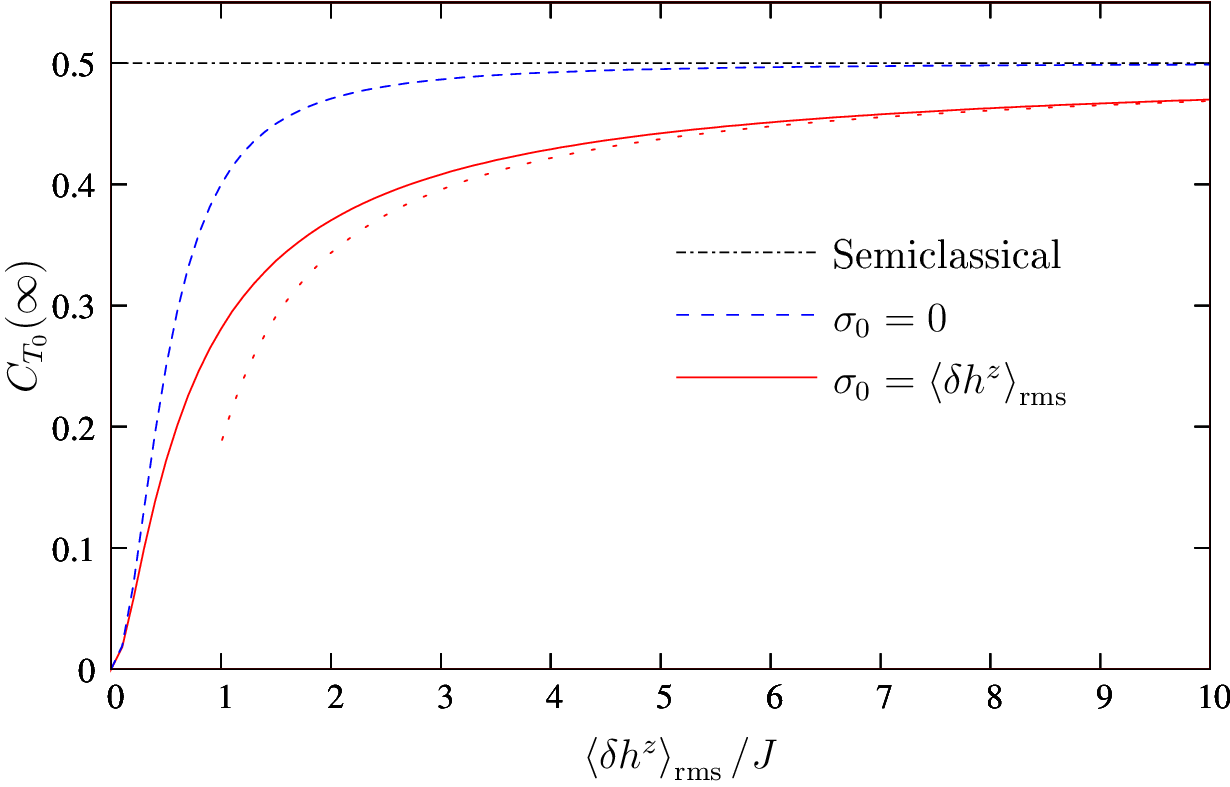}

\caption{\label{CorrelatorSaturationRatio}Saturation value of the triplet
occupation probability $C_{T_{0}}(\infty)=\overline{C_{n}}$ vs. $\left\langle \delta h^{z}\right\rangle _{\mathrm{rms}}/J$
when the nuclear spin system has been squeezed into an eigenstate
of $\delta h^{z}$ (dashed line) and when the state of the nuclear
spin system describes a Gaussian distribution of eigenvalues $\delta h_{n}^{z}$,
with mean $\overline{\delta h_{n}^{z}}=0$ and variance $\sigma_{0}^{2}=\overline{\left(\delta h_{n}^{z}\right)^{2}}$
(solid line). We also show the analytical asymptotics for $\sigma_{0}\gg J$,
given by Eq. (\ref{eq:CorrelatorSaturationLimits}) (dotted line)
and the semiclassical value ($C_{T_{0}}(\infty)=1/2$) (dash-dotted
line).}
\end{figure}

At times longer than the singlet-triplet decoherence time the interference
term vanishes, leaving $C_{T_{0}}(\infty)=\overline{C_{n}}$, which
depends only on the ratio $\delta h_{n}^{z}/J$, and could therefore
be used to trace-out the slow adiabatic dynamics $\delta h_{n}^{z}(t)$
of the nuclear spins, or to measure the exchange coupling $J$ when
the size of the hyperfine field fluctuations is known. We evaluate
$C_{T_{0}}(\infty)$ from \begin{equation}
C_{T_{0}}(\infty)=\overline{C_{n}}=\int_{-\infty}^{\infty}dxC(x)P_{\sigma_{0};0}(x).\label{eq:CSaturationIntegral}\end{equation}
In two limiting cases, we find the saturation value is given by (see
Appendix \ref{sec:Asymptotics})\begin{equation}
C_{T_{0}}(\infty)\sim\left\{ \begin{array}{c}
\frac{1}{2}-\sqrt{\frac{\pi}{2}}\frac{J}{4\sigma_{0}},\,\,\,\,\,\sigma_{0}\gg J,\\
2\left(\frac{\sigma_{0}}{J}\right)^{2},\,\,\,\,\,\sigma_{0}\ll J.\end{array}\right.\label{eq:CorrelatorSaturationLimits}\end{equation}
 We recover the semiclassical high-magnetic-field limit\cite{schulten:1978a}
$(C_{T_{0}}(\infty)=1/2)$ \emph{only} when the exchange $J$ is much
smaller than $\sigma_{0}$. Furthermore, due to the average over $\delta h_{n}^{z}$
eigenstates, the approach to the semiclassical value of $\frac{1}{2}$
is a slowly-varying (linear) function of the ratio $J/\sigma_{0}$,
in spite of the fact that $C_{n}\propto\left(J/\delta h_{n}^{z}\right)^{2}$
as $J\to0$. In Figure \ref{CorrelatorSaturationRatio} we plot the
correlator saturation value $C_{T_{0}}(\infty)$ as a function of
the ratio $\left\langle \delta h^{z}\right\rangle _{\mathrm{rms}}/J$
for a nuclear spin system described by a fixed eigenstate of $\delta h^{z}$
(i.e. $\rho_{I}=\ket{n}\bra{n}$), and for a nuclear spin system that
describes a Gaussian distribution of $\delta h^{z}$ eigenstates with
variance $\sigma_{0}^{2}=\overline{\left(\delta h_{n}^{z}\right)^{2}}=\left\langle \delta h^{z}\right\rangle _{\mathrm{rms}}^{2}$.
We also show the asymptotic expression for $\sigma_{0}\gg J$, as
given in Eq. (\ref{eq:CorrelatorSaturationLimits}). %

\begin{figure}
\includegraphics[%
  scale=0.65]{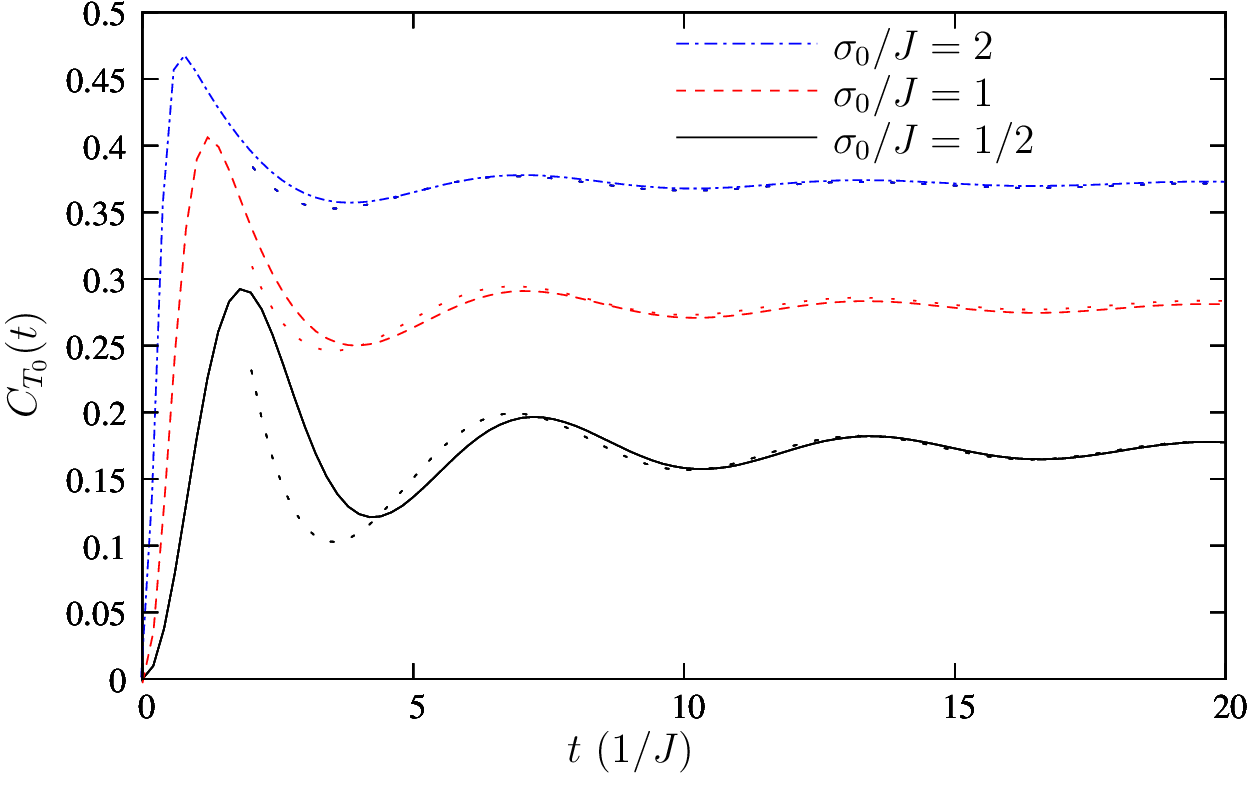}

\caption{\label{cap:TDepCorrelator}Decay of the correlator $C_{T_{0}}(t)$
evaluated by numerical integration of Eq. (\ref{eq:ComplexCT0}) for
three ratios of $\sigma_{0}/J$: $\sigma_{0}/J=2$ (dashed-dotted
line), $\sigma_{0}/J=1$ (dashed line), and $\sigma_{0}/J=1/2$ (solid
line). The analytical long-time asymptotic expressions from Eq. (\ref{eq:CT0LongTimes})
are shown as dotted lines. }
\end{figure}

Now we turn to the interference term $C_{T_{0}}^{\mathrm{int}}(t)$
given by Eq. (\ref{eq:ComplexCT0}), which can be evaluated explicitly
in several interesting limits. First, in the limiting case of vanishing
exchange $(J=0)$, we have $C(x)=\frac{1}{2}$ from (\ref{eq:Cofx}).
Direct integration of Eq. (\ref{eq:ComplexCT0}) then gives\begin{equation}
C_{T_{0}}^{\mathrm{int}}(t)=-\frac{1}{2}\exp\left(-\frac{t^{2}}{2t_{0}^{2}}\right),\, t_{0}=\frac{1}{2\sigma_{0}},\, J=0.\label{eq:ZeroJCorrelator}\end{equation}
 For zero exchange interaction, the correlator decays purely as a
Gaussian, with decoherence time $t_{0}=\frac{1}{2\sigma_{0}}\approx\frac{\sqrt{N}}{IA}$
for a typical asymmetric double quantum dot (see Appendix \ref{sec:Overhauserfieldestimates}).
However, for arbitrary nonzero exchange interaction $J\ne0$, we find
the asymptotic form of the correlator at long times is given by (see
Appendix \ref{sec:Asymptotics}):\begin{eqnarray}
C_{T_{0}}^{\mathrm{int}}\left(t\right) & \sim & -\frac{\cos\left(Jt+\frac{3\pi}{4}\right)}{4\sigma_{0}\sqrt{J}t^{3/2}},\label{eq:CT0LongTimes}\\
 &  & t\gg\max\left(\frac{1}{J},\frac{1}{2\sigma_{0}},\frac{J}{4\sigma_{0}^{2}}\right).\end{eqnarray}
 Thus, for arbitrarily small exchange interaction $J$, the asymptotic
decay law of the correlator is modified from the Gaussian behavior
of Eq. (\ref{eq:ZeroJCorrelator}) to a (much slower) power law ($\sim1/t^{3/2}$).
We also note that the long-time correlator has a universal phase shift
of $\frac{3\pi}{4}$, which is independent of any microscopic parameters.
Our calculation therefore provides an example of interesting non-Markovian
decay in an experimentally accessible system. Furthermore, the slow-down
of the asymptotic decay suggests that the exchange interaction can
be used to modify the \emph{form} of decay, in addition to the decoherence
time, through a narrowing of the distribution of eigenstates (see
the discussion following Eq. (\ref{eq:FrequencyExpansion}) below).
We have evaluated the full correlator $C_{T_{0}}(t)$ by numerical
integration of Eq. (\ref{eq:ComplexCT0}) and plotted the results
in Figure \ref{cap:TDepCorrelator} along with the analytical asymptotic
forms from (\ref{eq:CT0LongTimes}). %

We now investigate the relevant singlet-triplet correlator $C_{T_{0}}(t)$
in the limit of large exchange $J$. In this case, we have $x\lesssim\sigma_{0}\ll J$
for the typical $x$ contributing to the integral in Eq. (\ref{eq:ComplexCT0}).
Thus, we can expand the prefactor $C(x)$ and frequency term in the
integrand:\begin{eqnarray}
C(x) & \approx & 2\frac{x^{2}}{J^{2}},\label{eq:CxSmallx}\\
\sqrt{J^{2}+4x^{2}} & \approx & J+2\frac{x^{2}}{J}.\label{eq:FrequencyExpansion}\end{eqnarray}
 From Eq. (\ref{eq:FrequencyExpansion}) it is evident that the range
of frequencies that contribute to the correlator is suppressed by
$\sigma_{0}/J$ (increasing the exchange narrows the distribution
of eigenenergies that can contribute to decay). This narrowing of
the linewidth will increase the decoherence time. Moreover, the leading-order
$x^{2}$-dependence in (\ref{eq:FrequencyExpansion}) collaborates
with the Gaussian distribution of $\delta h^{z}$ eigenstates to induce
a power-law decay. With the approximations in Eqs. (\ref{eq:CxSmallx})
and (\ref{eq:FrequencyExpansion}), we find an expression for the
correlator that is valid for all times in the limit of large exchange
$J$ by direct evaluation of the integral in Eq. (\ref{eq:ComplexCT0}):\begin{eqnarray}
C_{T_{0}}^{\mathrm{int}}(t) & = & -2\left(\frac{\sigma_{0}}{J}\right)^{2}\frac{\cos\left(Jt+\frac{3}{2}\arctan\left(\frac{t}{t_{0}^{\prime}}\right)\right)}{\left(1+\left(\frac{t}{t_{0}^{\prime}}\right)^{2}\right)^{3/4}},\label{eq:LargeJCorrelator}\\
 &  & t_{0}^{\prime}=\frac{J}{4\sigma_{0}^{2}},\,\, J\gg\sigma_{0}.\end{eqnarray}
 There is a new timescale ($t_{0}^{\prime}=J/4\sigma_{0}^{2}$) that
appears for large $J$ due to dynamical narrowing; increasing the
exchange $J$ results in rapid precession of the pseudospin $\pmb{\tau}$
about the $z$-axis, which makes transverse fluctuations along $\tau^{x}$
due to $\delta h^{z}$ progressively unimportant. Explicitly, we have
$t_{0}^{\prime}\approx JN/4A^{2}\gg\sqrt{N}/A$ for $J\gg\sigma_{0}\approx A/\sqrt{N}$.

\begin{figure}
\includegraphics[%
  scale=0.65]{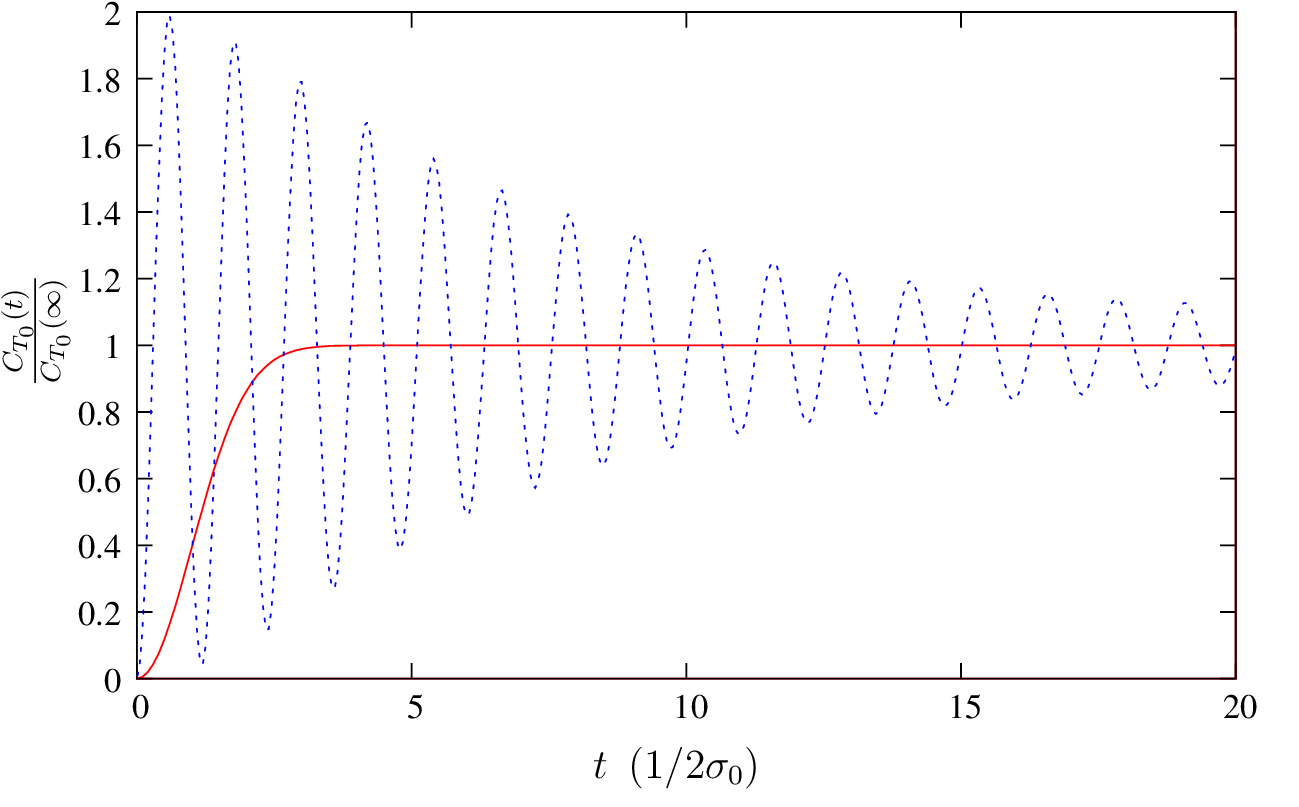}

\caption{\label{cap:LargeJSmallJCT0}The correlator $C_{T_{0}}(t)/C_{T_{0}}(\infty)$
shows a rapid Gaussian decay when $J=0$ (solid line, from Eq. (\ref{eq:ZeroJCorrelator})),
but has a much slower power-law decay $\sim1/t^{3/2}$ for large exchange
$J=10\sigma_{0}\gg\sigma_{0}$ (dotted line, from Eq. (\ref{eq:LargeJCorrelator})). }
\end{figure}

Eq. (\ref{eq:LargeJCorrelator}) provides a potentially useful means
of extracting the relevant microscopic parameters from an experiment.
$J$ and $\sigma_{0}$ can be determined independent of each other
exclusively from a measurement of the oscillation frequency and phase
shift of $C_{T_{0}}^{\mathrm{int}}(t)$. In particular, any loss of
oscillation amplitude (visibility) due to systematic error in the
experiment can be ignored for the purposes of finding $\sigma_{0}$
and $J$. The loss in visibility can then be quantified by comparison
with the amplitude expected from Eq. (\ref{eq:LargeJCorrelator}).
We illustrate the two types of decay that occur for large and small
$J$ in Figure \ref{cap:LargeJSmallJCT0}.

\subsection{\label{sub:Deltahnonzero}Inhomogeneous polarization, $\overline{\delta h_{n}^{z}}\ne0$}

It is possible that a nonequilibrium inhomogeneous average polarization
could be generated in the nuclear spin system, in which case $\overline{\delta h_{n}^{z}}\ne0$.
Pumping of nuclear spin polarization occurs naturally, for example,
at donor impurities in GaAs during electron spin resonance (ESR),
resulting in a shift of the ESR resonance condition.\cite{seck:1997a}
It is therefore important to investigate the effects of a nonzero
average Overhauser field inhomogeneity on the decay law and timescale
of the singlet-triplet correlator. In this subsection we generalize
our previous results for the case $\overline{\delta h_{n}^{z}}\ne0$.

We set the mean Overhauser field inhomogeneity to $\overline{\delta h_{n}^{z}}=x_{0}$,
in which case the complex singlet-triplet interference term is given
by \begin{eqnarray}
\tilde{C}_{T_{0}}^{\mathrm{int}}(t) & = & -\int_{-\infty}^{\infty}dxC(x)P_{\sigma_{0};x_{0}}(x)e^{it\sqrt{J^{2}+4x^{2}}}.\label{eq:ComplexCT0x0nonzero}\end{eqnarray}
When the mean value of the Overhauser field inhomogeneity $x_{0}$
is much larger than the fluctuations $\sigma_{0}$ ($x_{0}\gg\sigma_{0}$),
we approximate $C(x)\approx C(x_{0})$ and expand the frequency term
$\sqrt{J^{2}+4x^{2}}=\omega_{0}+\frac{4x_{0}}{\omega_{0}}(x-x_{0})+\frac{2J^{2}}{\omega_{0}^{3}}(x-x_{0})^{2}+\cdots$,
where $\omega_{0}=\sqrt{J^{2}+4x_{0}^{2}}$. We retain only linear
order in $x-x_{0}$ for the frequency term, which is strictly valid
for times $t\ll(J^{2}+4x_{0}^{2})^{3/2}/2J^{2}\sigma_{0}^{2}$. This
time estimate is found by replacing $\left(x-x_{0}\right)^{2}\approx\sigma_{0}^{2}$
in the quadratic term and demanding that the quadratic term multiplied
by time be much less than one. In this limit, the correlator and range
of validity are then \begin{eqnarray}
C_{T_{0}}^{\mathrm{int}}(t) & = & -\frac{2x_{0}^{2}}{\omega_{0}^{2}}e^{-\frac{1}{2}\left(\frac{t}{t_{0}^{\prime\prime}}\right)^{2}}\cos\left(\omega_{0}t\right),\label{eq:CintLargex0}\\
 &  & t_{0}^{\prime\prime}=\frac{\omega_{0}}{4x_{0}\sigma_{0}},\,\,\,\,\,\omega_{0}=\sqrt{J^{2}+4x_{0}^{2}},\\
 &  & x_{0}\gg\sigma_{0},\,\,\,\,\, t\ll\frac{\left(J^{2}+4x_{0}^{2}\right)^{3/2}}{2J^{2}\sigma_{0}^{2}}.\label{eq:CintLargex0Validity}\end{eqnarray}
 This expression is valid for any value of the exchange $J$, up to
the timescale indicated.

In contrast with the previous result for $x_{0}=0$, from Eq. (\ref{eq:CintLargex0})
we find that the long-time saturation value of the correlator deviates
from the semiclassical result ($C_{T_{0}}(\infty)=-C_{T_{0}}^{\mathrm{int}}(0)=1/2$)
by an amount that is quadratic in the exchange $J$ for $J\ll x_{0}$:\begin{equation}
C_{T_{0}}(\infty)=C_{T_{0}}^{\mathrm{int}}(0)\sim\left\{ \begin{array}{c}
\frac{1}{2}-\frac{1}{8}\left(\frac{J}{x_{0}}\right)^{2},\,\, J\ll x_{0},\\
2\left(\frac{x_{0}}{J}\right)^{2},\,\, J\gg x_{0}.\end{array}\right.,\,\,\,\,\, x_{0}\gg\sigma_{0}.\end{equation}

In the limit of large exchange, $J\gg\max\left(\sigma_{0},x_{0}\right)$,
we can once again apply the approximations given in Eqs. (\ref{eq:CxSmallx})
and (\ref{eq:FrequencyExpansion}). Using these approximations in
Eq. (\ref{eq:ComplexCT0x0nonzero}) and integrating then gives

\begin{widetext}\begin{eqnarray}
\tilde{C}_{T_{0}}^{\mathrm{int}}(t) & = & -2\left(\frac{\sigma_{0}}{J}\right)^{2}\xi^{3}(t)\left(1+\left(\frac{x_{0}}{\sigma_{0}}\right)^{2}\xi^{2}(t)\right)\exp\left\{ iJt-\frac{x_{0}^{2}}{2\sigma_{0}^{2}}\left(1-\xi^{2}(t)\right)\right\} ,\label{eq:CintNonzerox0}\\
 &  & \xi(t)=\left(1-i\frac{t}{t_{0}^{\prime}}\right)^{-1/2},\,\, t_{0}^{\prime}=\frac{J}{4\sigma_{0}^{2}},\,\,\,\,\, J\gg\max(x_{0},\sigma_{0}),\,\, t\ll\frac{J^{3}}{2\max\left(x_{0}^{4},\sigma_{0}^{4}\right)}.\label{eq:CintNonzerox0validity}\end{eqnarray}
 \end{widetext} We have found the limit on the time range of validity
in Eq. (\ref{eq:CintNonzerox0validity}) using the same estimate that
was used for Eqs. (\ref{eq:CintLargex0}-\ref{eq:CintLargex0Validity}).
At short times, $t\ll t_{0}^{\prime}=J/4\sigma_{0}^{2}$, we expand
$\xi^{2}(t)\approx1+i\frac{t}{t_{0}^{\prime}}-\left(\frac{t}{t_{0}^{\prime}}\right)^{2}$
and find that this function decays initially as a Gaussian with timescale
$t_{0}^{\prime\prime}\approx J/4x_{0}\sigma_{0}$:\begin{eqnarray}
C_{T_{0}}^{\mathrm{int}}(t) & \sim & -2\frac{\sigma_{0}^{2}+x_{0}^{2}}{J^{2}}e^{-\frac{1}{2}\left(\frac{t}{t_{0}^{\prime\prime}}\right)^{2}}\cos\left(\omega_{0}^{\prime}t\right),\\
 &  & t_{0}^{\prime\prime}\approx\frac{J}{4x_{0}\sigma_{0}},\,\,\,\,\,\omega_{0}^{\prime}=J+\frac{2x_{0}^{2}}{J},\\
 &  & t\ll t_{0}^{\prime}=\frac{J}{4\sigma_{0}^{2}},\,\,\,\,\, J\gg\max(x_{0},\sigma_{0}).\end{eqnarray}
This agrees with the result in Eq. (\ref{eq:CintLargex0}) when $J\gg x_{0}\gg\sigma_{0}$. 

For sufficiently large exchange $J$, the expression given by Eq.
(\ref{eq:CintNonzerox0}) is valid for times longer than the previous
expression, given by Eq. (\ref{eq:CintLargex0}). We perform an asymptotic
expansion of Eq. (\ref{eq:CintNonzerox0}) for long times using $\xi(t\gg t_{0}^{\prime})\sim e^{i\pi/4}\sqrt{t_{0}^{\prime}/t}$.
This gives\begin{eqnarray}
C_{T_{0}}^{\mathrm{int}}(t) & \sim & -\frac{e^{-x_{0}^{2}/2\sigma_{0}^{2}}\cos(Jt+\frac{3\pi}{4})}{4\sigma_{0}\sqrt{J}t^{3/2}},\label{eq:CintAsympx0nonzero}\\
 &  & t\gg t_{0}^{\prime}=\frac{J}{4\sigma_{0}^{2}},\,\,\,\,\, J\gg\max(x_{0},\sigma_{0}).\end{eqnarray}
As in the case of $x_{0}=0$, the long-time asymptotics of Eq. (\ref{eq:CintNonzerox0})
once again give a power law $\sim1/t^{3/2}$, although the amplitude
of the long-time decay is exponentially suppressed in the ratio $x_{0}^{2}/\sigma_{0}^{2}$.
When $x_{0}=0$, Eq. (\ref{eq:CintAsympx0nonzero}) recovers the previous
result, given in Eq. (\ref{eq:CT0LongTimes}).

\subsection{Reducing decoherence}

The results of this section suggest a general strategy for increasing
the amplitude of coherent oscillations between the singlet $\ket{S}$
and triplet $\ket{T_{0}}$ states, and for weakening the form of decay.
To avoid a rapid Gaussian decay with a timescale $t_{0}^{\prime\prime}=J/4x_{0}\sigma_{0}$,
the mean Overhauser field inhomogeneity should be made smaller than
the fluctuations ($\overline{\delta h_{n}^{z}}=x_{0}\lesssim\sigma_{0}$)
and the exchange $J$ should be made larger than $x_{0}$ and $\sigma_{0}$
($J\gg\max(x_{0},\sigma_{0})$). Explicitly, the ideal condition for
slow and weak (power-law) decay can be written as\begin{equation}
J\gg\sigma_{0}\gtrsim x_{0}.\label{eq:SlowDecayCondition}\end{equation}
 The condition in Eq. (\ref{eq:SlowDecayCondition}) can be achieved
equally well by increasing the exchange coupling $J$ for fixed hyperfine
fluctuations $\sigma_{0}$ or by reducing the fluctuations $\sigma_{0}$
through state squeezing or by making the double-dot confining potential
more symmetric (see Appendix \ref{sec:Overhauserfieldestimates}).

\section{\label{sec:STplus}Dynamics in the subspace of $\ket{S}$ and $\ket{T_{+}}$}

\begin{figure}
\includegraphics[%
  scale=0.65]{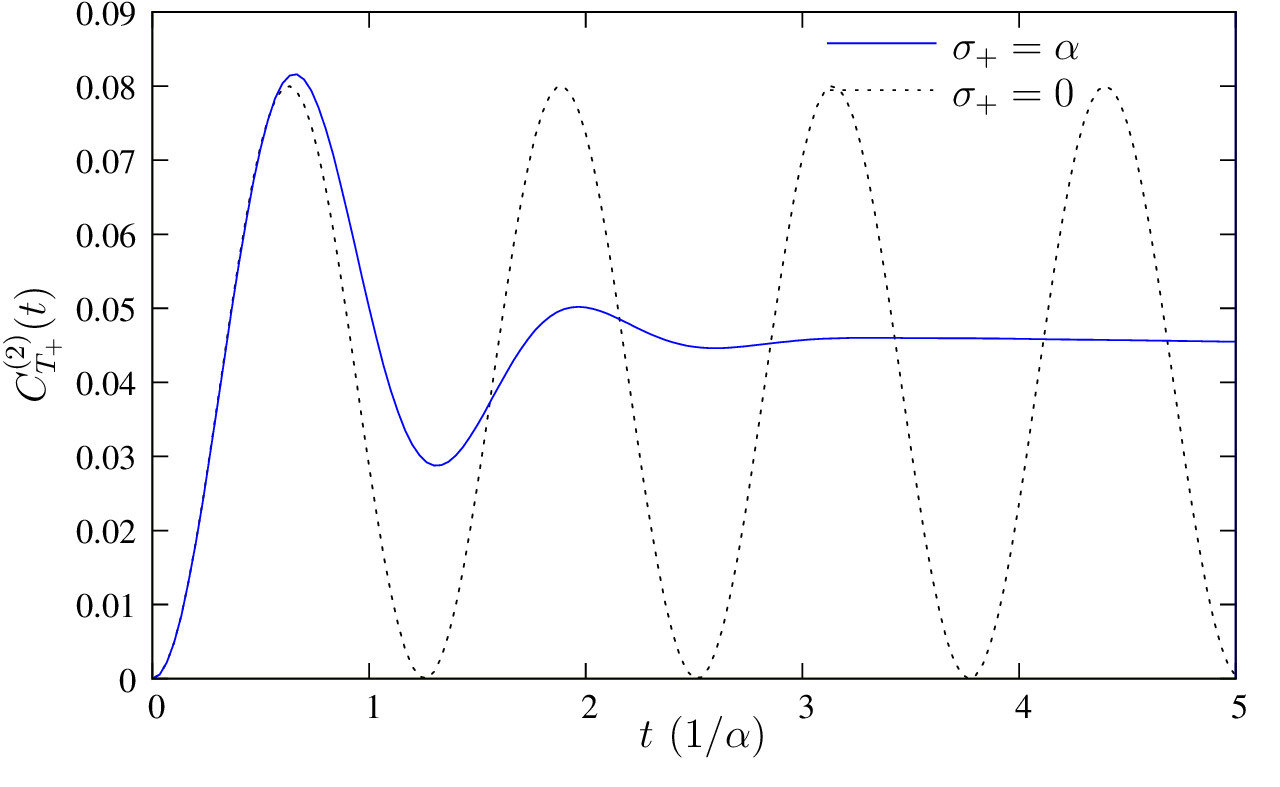}

\caption{\label{cap:CTplusCorrelator}Decay of the correlator $C_{T_{+}}(t)$
in two cases. A typical unprepared initial state, where the nuclear
spin system is in a superposition of $h^{z}$ eigenstates, results
in a Gaussian decay (solid line, from Eq. (\ref{eq:CTplus})). If
the nuclear spin state is squeezed into an $h^{z}$ eigenstate, there
is no decay, only coherent oscillations (dotted line, from Eq. (\ref{eq:PerturbationCTplus})).
For this plot we have used $\Delta=5\alpha$.}
\end{figure}

We now consider the case when the Zeeman energy of the $S^{z}=1$
triplet state approximately compensates the exchange ($\left|\Delta\right|\ll J$,
where $\Delta=\epsilon_{z}+J$). In addition, we assume the exchange
is much larger than the nuclear field energy scales $J\gg\mathrm{max}\left\{ \left\langle \mathbf{\delta h}\right\rangle _{\mathrm{rms}},\left\langle \mathbf{h}\right\rangle _{\mathrm{rms}}\right\} $.
Under these conditions, we consider the dynamics in a subspace formed
by the singlet $\ket{S}\to\ket{\tau^{z}=-1}$ and the $S^{z}=1$ triplet
state $\ket{T_{+}}\to\ket{\tau^{z}=+1}$, governed by the Hamiltonian
(to zeroth order in $1/J$, see Appendix \ref{sec:Effective-Hamiltonians}):\begin{equation}
H_{+}=\frac{1}{2}\left(\Delta+h^{z}\right)\left(1+\tau^{z}\right)-\frac{1}{\sqrt{2}}\left(\delta h^{-}\tau^{+}+\mathrm{H.c.}\right).\end{equation}
 Here, $\delta h^{\pm}=\delta h^{x}\pm i\delta h^{y}$ and $\tau^{\pm}=\frac{1}{2}\left(\tau^{x}\pm i\tau^{y}\right)$.
The $\ket{T_{+}}$ probability at time $t>0$ is\begin{equation}
C_{T_{+}}(t)=\sum_{n,n^{\prime}}\rho_{I}(n)\left|\bra{n^{\prime}}\otimes\bra{T_{+}}e^{-iH_{+}t}\ket{S}\otimes\ket{n}\right|^{2}.\end{equation}
 This case is essentially different from the previous one, since the
eigenstates of $H_{+}$ are no longer simply product states of electron
and nuclear spin, implying a back-action of the electron on the nuclear
system. Nevertheless, when $\left\langle h^{z}+\Delta\right\rangle _{\mathrm{rms}}\gg\left\langle \mathbf{\delta h^{\pm}}\right\rangle _{\mathrm{rms}}$,
we can evaluate the correlator in standard time-dependent perturbation
theory to leading order in the term $V=-\frac{1}{\sqrt{2}}\left(\tau^{+}\delta h^{-}+\tau^{-}\delta h^{+}\right)$.
Neglecting corrections of order $h_{n}^{z}/\Delta\ll1$, this gives\begin{equation}
C_{T_{+}}^{(2)}(t)\approx\overline{\frac{\alpha_{n}^{2}}{\Delta^{2}}\left(1-\cos\left(\left[\left[h^{z}\right]_{n}+\Delta\right]t\right)\right)},\label{eq:PerturbationCTplus}\end{equation}
where $\alpha_{n}=\sum_{n^{\prime}}\left|\bra{n^{\prime}}\delta h^{-}\ket{n}\right|^{2}$,
and $\ket{n}$ is now an eigenstate of the operator $h^{z}$ with
eigenvalue $\left[h^{z}\right]_{n}$. To estimate the size of $\alpha_{n}$,
we assume identical completely decoupled dots and nuclear polarization
$p\ll1$, which gives $\alpha_{n}^{2}\approx\frac{1}{2}I(I+1)\sum_{k}A_{k}^{2}$,
where $A_{k}$ is the hyperfine coupling constant to the nuclear spin
at lattice site $k$ (with total nuclear spin $I$) and the sum $\sum_{k}$
runs over all lattice sites in one of the dots. We estimate the typical
size of $\alpha_{n}$ with the replacements $A_{k}\to\frac{A}{N},\,\sum_{k}\to N$,
which gives $\alpha_{n}\approx\alpha/\sqrt{2}=\sqrt{\frac{I(I+1)}{2N}}A$,
where $N$ characterizes the number of nuclear spins within the dot
envelope wavefunction. If we assume the nuclear spin state is described
by a continuous Gaussian distribution of $h^{z}$ eigenstates with
mean $\overline{h_{n}^{z}}=0$ and variance $\sigma_{+}^{2}$, we
find \begin{equation}
C_{T_{+}}^{(2)}(t)\approx\frac{1}{2}\left(\frac{\alpha}{\Delta}\right)^{2}\left(1-e^{-t^{2}/2t_{+}^{2}}\cos\left(\Delta t\right)\right),\,\,\,\, t_{+}=\frac{1}{2\sigma_{+}}.\label{eq:CTplus}\end{equation}
 Thus, if we ignore any possibility for recurrence, the distribution
of $h^{z}$ eigenstates will lead to Gaussian decay of the two-electron
spin state, as is the case for a single electron.\cite{schliemann:2002a,coish:2004a}
However, as in the case of a single electron, this decay can be reduced
or eliminated altogether by narrowing the distribution of $h^{z}$
eigenstates $\ket{n}$ through measurement (squeezing the nuclear
spin state).\cite{coish:2004a} We show these two cases (with and
without squeezing of the nuclear state) in Figure \ref{cap:CTplusCorrelator}.%

\section{\label{sec:Orbital-dephasing}Singlet-triplet decoherence due to
orbital dephasing}

To this point we have neglected dephasing of the singlet $\ket{S}$
and triplet $\ket{T_{j}}$ $(j=0,+)$ states due to coupling in the
orbital sector. The effective Hamiltonian description ignores the
different character of the orbital states for singlet and triplet,
and so it is tempting to assume that orbital dephasing is unimportant
where the effective Hamiltonian is valid. However, the singlet and
triplet do have different orbital states which can, in general, couple
differently to the environment through the charge degree of freedom,
and therefore acquire different phases. Examples of such environmental
influences are charge fluctuators or measurement devices, such as
quantum point contacts used for charge readout.\cite{engel:2004a,elzerman:2004a}
Here we briefly step away from the effective Hamiltonians derived
in Appendix \ref{sec:Effective-Hamiltonians} to give a physical picture
of the effects of orbital dephasing in terms of the true double-dot
wavefunctions. We then return to the effective Hamiltonian picture
in order to give a more general estimate of the effects of orbital
dephasing on singlet-triplet decoherence for a two-electron double
dot. 

We consider a double quantum dot containing a fixed (quantized) number
of electrons $N$. Within the far-field approximation, the double-dot
charge distribution couples to the environment first through a monopole,
and then a dipole term. Since the charge on the double dot is quantized,
the monopole term gives an equal contribution for both the singlet
and triplet wavefunctions. The leading interaction that can distinguish
singlet from triplet is the electric dipole term: \begin{equation}
V_{\mathrm{orb}}(t)\approx-\mathbf{p}_{N}\cdot\mathbf{E}(t).\end{equation}
 Here, $\mathbf{p}_{N}$ is the electric dipole moment operator for
the charge distribution in a double dot containing $N$ electrons
and $\mathbf{E}(t)$ is a fluctuating electric field due to the surrounding
environment, which we model by a Gaussian random process. For a double
quantum dot with well-localized single-particle eigenstates we denote
the charge states by $\ket{(n,m)}$, indicating that the double-dot
has $n$ electrons in dot $1$ and $m$ electrons in dot $2$, where
$n+m=N$. If the double dot contains only a single electron $(N=1)$,
the environment can distinguish the two localized states through the
difference in the dipole moment operator, which has the size $\left|\Delta\mathbf{p}_{1}\right|=\left|\bra{(1,0)}\mathbf{p}_{1}\ket{(1,0)}-\bra{(0,1)}\mathbf{p}_{1}\ket{(0,1)}\right|\approx2\left|e\right|a$,
where $e$ is the electron charge and $2a$ is the inter-dot spacing.
When $N=2$, for highly-localized states, only the states with double-occupancy
($\ket{(0,2)}$ and $\ket{(2,0)}$) contribute to the dipole moment.
If the typical hyperfine energy scale is much smaller than the detuning
from resonance $\delta$ of the $\ket{(1,1)}$ and $\ket{(0,2)}$
states ($\max\left(\left\langle \delta\mathbf{h}\right\rangle _{\mathrm{rms}},\left\langle \mathbf{h}\right\rangle _{\mathrm{rms}}\right)\ll\delta$),
only the $\ket{(1,1)}$ singlet state (not the triplets) will mix
with the doubly-occupied states, so the singlet and triplet states
will be energetically distinguishable through $\left|\Delta\mathbf{p}_{2}\right|=\left|\bra{S}\mathbf{p}_{2}\ket{S}\right|\approx2\left|e\right|a\left|P_{(0,2)}-P_{(2,0)}\right|\lesssim2\left|e\right|aD$,
where $P_{(0,2)}$ $\left(P_{(2,0)}\right)$ is the probability to
find the singlet $\ket{S}$ in the $\ket{(0,2)}$ $(\ket{(2,0)})$
state and $D=P_{(0,2)}+P_{(2,0)}$ is the double occupancy. In this
discussion, we assume that the exchange is much larger than the hyperfine
energy scales, $J\gg\max\left(\left\langle \mathbf{h}\right\rangle _{\mathrm{rms}},\left\langle \delta\mathbf{h}\right\rangle _{\mathrm{rms}}\right)$,
so that the singlet and triplet states are good approximates for the
true two-electron eigenstates.

For weak coupling to the environment, and assuming the environment
correlation time is much less than the orbital dephasing time $t_{\phi}^{(N)}$,
we can apply standard techniques to determine the dephasing time for
a two-level system described by the Bloch equations.\cite{blum:1981a}
We find that the fluctuations in $\mathbf{E}(t)$ lead to exponential
dephasing with the rate $1/t_{\phi}^{(N)}=\frac{1}{4}\left|\Delta\mathbf{p}_{N}\right|^{2}\int_{-\infty}^{\infty}dtE(t)E(0)$,
where the scalar $E(t)$ is the component of $\mathbf{E}(t)$ along
$\Delta\mathbf{p}_{N}$ and we assume $\lim_{t\to\infty}\frac{1}{t}\int_{0}^{t}dt^{\prime}E(t^{\prime})=0$.
Assuming equivalent environments for the single-particle and two-particle
cases, the ratio of the single-particle to two-particle dephasing
times is then\begin{equation}
\frac{t_{\phi}^{(1)}}{t_{\phi}^{(2)}}=\left|\frac{\Delta\mathbf{p}_{2}}{\Delta\mathbf{p}_{1}}\right|^{2}\lesssim D^{2}.\label{eq:TphiDoubleOccupancyRelation}\end{equation}
 The single-electron orbital dephasing rate has been measured to be
$t_{\phi}^{(1)}\approx1\,\mathrm{ns}$\cite{hayashi:2003a} and $t_{\phi}^{(1)}\approx400\,\mathrm{ps}$\cite{petta:2004b}
in different gated double quantum dots. If the hyperfine interaction
(which becomes important on the timescale $t\gtrsim5\,\mathrm{ns}$)
is to provide the major source of decoherence in these two-electron
structures, we therefore require $t_{\phi}^{(2)}\gg t_{\phi}^{(1)}$.
This condition can be achieved by ensuring a small double occupancy
$D\ll1$ of the singlet state. When the inter-dot tunnel coupling
$t_{12}$ is much less than the detuning from resonance $\delta$
($t_{12}\ll\delta\ll U+U^{\prime}$, with on-site and nearest-neighbor
charging energies $U$ and $U^{\prime}$, respectively -- see Appendix
\ref{sec:Effective-Hamiltonians}) we find the double-occupancy of
$\ket{S}$ in perturbation theory is\begin{equation}
D\approx2\left(\frac{t_{12}}{\delta}\right)^{2}\ll1.\label{eq:PerturbativeDoubleOccupancy}\end{equation}
Even in this regime, orbital dephasing may become the limiting timescale
for singlet-triplet decoherence after the removal of hyperfine-induced
decoherence by spin echo. A detailed analysis of the double-occupancy
and its relation to the concurrence (an entanglement measure) for
a symmetric double dot can be found in Refs. \onlinecite{golovach:2003a,golovach:2004b}. 

With this physical picture in mind, we can generalize the above results
to the case when the electrons experience fluctuations due to any
time-dependent classical fields. In particular, if the separation
in single-particle energy eigenstates for $N=1$ is $\epsilon+\delta\epsilon(t)$,
where $\delta\epsilon(t)$ fluctuates randomly with amplitude $\delta\epsilon$,
and similarly, if for $N=2$ the singlet and triplet levels are separated
by an exchange $J+\delta J(t)$, where $\delta J(t)$ has amplitude
$\delta J$, we find \begin{equation}
\frac{t_{\phi}^{(1)}}{t_{\phi}^{(2)}}=\left|\frac{\delta J}{\delta\epsilon}\right|^{2}.\label{eq:TPhiFluctuationsRelation}\end{equation}
 From this expression we conclude that the optimal operating point
of the double dot is where the slope of $J$ vs. $\epsilon$ vanishes,
i.e., $\delta J/\delta\epsilon=0$. At this optimal point, $t_{\phi}^{(2)}\to\infty$,
within the approximations we have made. Eq. (\ref{eq:TPhiFluctuationsRelation})
is valid for weak coupling to the environment (i.e. $\delta J\ll J$
and $\delta\epsilon\ll\epsilon$), and when the environment correlation
time is small compared to the dephasing times. If, for example, we
take $J\approx2t_{12}^{2}/\delta$ for $U+U^{\prime}\gg\delta\gg t_{12}$
from Eq. (\ref{eq:ExchangeLargedelta}) and if $\delta\epsilon$ corresponds
to fluctuations in the single-particle charging energy difference
($\epsilon\sim(V_{g1}-V_{g2})\sim\delta$ from Eq. (\ref{eq:delta2definition})),
we find $t_{\phi}^{(1)}/t_{\phi}^{(2)}\approx4t_{12}^{4}/\delta^{4}$,
in agreement with Eqs. (\ref{eq:TphiDoubleOccupancyRelation}) and
(\ref{eq:PerturbativeDoubleOccupancy}). In particular, the hyperfine-dominated
singlet-triplet decoherence becomes visible when $t_{\phi}^{(2)}\gg t_{0}^{\prime},\, t_{0}^{\prime\prime}\gg t_{0},\, t_{+}$.
This regime is achievable by choosing $\delta\gg t_{12}$, but still
$J\approx2t_{12}^{2}/\delta\gg\sigma_{0}$, since $t_{\phi}^{(2)}$
is a much stronger function of $\delta$ than $t_{0}^{\prime},\, t_{0}^{\prime\prime}$.
That is, the two-particle dephasing time scales like $t_{\phi}^{(2)}\sim\delta^{4}$,
but the typical hyperfine-induced decay times scale like $t_{0}^{\prime},t_{0}^{\prime\prime}\sim J\sim1/\delta$.
On the other hand, when $t_{12}\approx\delta$, we have $\left|\delta J/\delta\epsilon\right|\sim O(1)$,
which gives $t_{\phi}^{(2)}\sim t_{\phi}^{(1)}$, and thus a very
short singlet-triplet decoherence time ($\approx1\,\mathrm{ns}$),
which is dominated by orbital dephasing.

\section{\label{sec:Conclusions}Conclusions}

We have shown that a fully quantum mechanical solution is possible
for the dynamics of a two-electron system interacting with an environment
of nuclear spins under an applied magnetic field. Our solution shows
that the singlet-triplet correlators $C_{T_{0}}(t)$ and $C_{T_{+}}(t)$
will decay due to the quantum distribution of the nuclear spin system,
even for a nuclear system that is static. We have found that the asymptotic
behavior of $C_{T_{0}}(t)$ undergoes a transition from Gaussian to
power-law ($\sim1/t^{3/2}$) when the Heisenberg exchange coupling
$J$ becomes nonzero, and acquires a universal phase shift of $3\pi/4$.
The oscillation frequency and phase shift as a function of time can
be used to determine the exchange and Overhauser field fluctuations.
We have also investigated the effects of an inhomogeneous polarization
on $C_{T_{0}}(t)$, and have suggested a general strategy for reducing
decoherence in this system. Finally, we have discussed orbital dephasing
and its effect on singlet-triplet decoherence.

\begin{acknowledgments}
We thank G. Burkard, J. C. Egues, J. A. Folk, V. N. Golovach, D. Klauser,
F. H. L. Koppens, J. Lehmann, C. M. Marcus, and J. R. Petta for useful
discussions. We acknowledge financial support from the Swiss NSF,
the NCCR nanoscience, EU RTN Spintronics, EU RTN QuEMolNa, EU NoE
MAGMANet, DARPA, ARO, ONR, and NSERC of Canada.
\end{acknowledgments}
\appendix

\section{\label{sec:Overhauserfieldestimates}Estimating the Overhauser field}

In this appendix we estimate the size of the Overhauser field inhomogeneity
for a typical double quantum dot, and show that this quantity depends,
in a sensitive way, on the form of the orbital wavefunctions.

As in the main text, we take the average Overhauser field and the
Overhauser field inhomogeneity to be $\mathbf{h}=\frac{1}{2}\left(\mathbf{h}_{1}+\mathbf{h}_{2}\right)$
and $\delta\mathbf{h}=\frac{1}{2}\left(\mathbf{h}_{1}-\mathbf{h}_{2}\right)$
respectively, where $\mathbf{h}_{l}=Av\sum_{k}\left|\psi_{0}^{l}(\mathbf{r}_{k})\right|^{2}\mathbf{I}_{k}$,
and $\psi_{0}^{l}(\mathbf{r})$ is orbital eigenstate $l$ in the
double quantum dot. In the presence of tunneling, the eigenstates
of a symmetric double quantum dot will be well-described\cite{burkard:1999a,golovach:2004b}
by the symmetric and antisymmetric linear combination of dot-localized
states $\phi_{l}(\mathbf{r}),\, l=1,2$: $\psi_{0}^{1,2}(\mathrm{r})=\frac{1}{\sqrt{2}}\left(\phi_{1}(\mathrm{r})\pm\phi_{2}(\mathrm{r})\right)$.
In this case, we find \begin{equation}
\left\langle \delta\mathbf{h}\right\rangle _{\mathrm{rms}}=Av\left\langle \sum_{k}\mathrm{Re}\left[\phi_{1}^{*}(\mathbf{r}_{k})\phi_{2}(\mathbf{r}_{k})\right]\mathbf{I}_{k}\right\rangle _{\mathrm{rms}}.\end{equation}
 We take $\left\langle \frac{1}{N}\sum_{k}\mathbf{I}_{k}\right\rangle _{\mathrm{rms}}\approx\sqrt{I(I+1)/N}$
to be the r.m.s. value for a system of $N$ nuclear spins with uniform
polarization $p\ll1$. Changing the sum to an integral according to
$v\sum_{k}\to\int d^{3}r$ then gives\begin{equation}
\left\langle \delta\mathbf{h}\right\rangle _{\mathrm{rms}}\approx\gamma\sqrt{\frac{I(I+1)}{N}}A=\gamma\alpha,\label{eq:deltahoverlap}\end{equation}
 where $\gamma=\int d^{3}r\,\mathrm{Re}\left[\phi_{1}^{*}(\mathbf{r})\phi_{2}(\mathbf{r})\right]$
is the overlap of the localized orbital dot states and we have introduced
the energy scale $\alpha=\sqrt{I(I+1)}A/\sqrt{N}$. The result in
Eq. (\ref{eq:deltahoverlap}) suggests that the Overhauser field inhomogeneity
can be drastically reduced in a symmetric double quantum dot simply
by separating the two dots, reducing the wavefunction overlap. If,
however, the double dot is sufficiently asymmetric, the correct orbital
eigenstates will be well-described by localized states $\psi_{0}^{l}(\mathrm{r})=\phi_{l}(\mathrm{r}),\,\, l=1,2$,
(with overlap $\gamma\ll1$), in which case we find \begin{equation}
\left\langle \delta\mathbf{h}\right\rangle _{\mathrm{rms}}\approx\sqrt{\frac{I(I+1)}{N}}A=\alpha.\label{eq:deltahestimatelocalizedstates}\end{equation}
 Thus, great care should be taken in determining $\left\langle \delta\mathbf{h}\right\rangle _{\mathrm{rms}}$
based on microscopic parameters. In particular, for a symmetric double
quantum dot, the overlap $\gamma$ must also be known to determine
$\left\langle \delta\mathbf{h}\right\rangle _{\mathrm{rms}}$ based
on $N$.

In contrast, for the total Overhauser operator $\mathbf{h}$, in both
of the above cases ($\psi_{0}^{1,2}(\mathrm{r})=\frac{1}{\sqrt{2}}\left(\phi_{1}(\mathrm{r})\pm\phi_{2}(\mathrm{r})\right)$
or $\psi_{0}^{l}(\mathrm{r})=\phi_{l}(\mathrm{r}),\,\, l=1,2$), we
find \begin{equation}
\left\langle \mathbf{h}\right\rangle _{\mathrm{rms}}\approx\sqrt{\frac{I(I+1)}{N}}A=\alpha.\end{equation}

\section{\label{sec:Effective-Hamiltonians}Effective Hamiltonians for two-electron
states in a double quantum dot}

In this appendix we derive effective Hamiltonians for a two-electron
system interacting with nuclear spins in a double quantum dot via
the contact hyperfine interaction. 

We begin from the two-electron Hamiltonian in second-quantized form,
\begin{equation}
H=H_{SP}+H_{C}+H_{T}+H_{Z}+H_{\mathrm{hf}},\end{equation}
 where $H_{SP}$ describes the single-particle charging energy, $H_{C}$
models the Coulomb interaction between electrons in the double dot,
$H_{T}$ describes tunneling between dot orbital states, $H_{Z}$
gives the electron Zeeman energy (we neglect the nuclear Zeeman energy,
which is smaller by the ratio of nuclear to Bohr magneton: $\mu_{N}/\mu_{B}\sim10^{-3}$)
and $H_{\mathrm{hf}}$ describes the Fermi contact hyperfine interaction
between electrons on the double dot and nuclei in the surrounding
lattice. Explicitly, these terms are given by\begin{eqnarray}
H_{SP} & = & \sum_{l\sigma}V_{gl}n_{l\sigma};\,\,\,\,\, n_{l\sigma}=d_{l\sigma}^{\dagger}d_{l\sigma},\\
H_{C} & = & U\sum_{l}n_{l\uparrow}n_{l\downarrow}+U^{\prime}(n_{1\uparrow}+n_{1\downarrow})(n_{2\uparrow}+n_{2\downarrow}),\\
H_{T} & = & t_{12}\sum_{\sigma}\left(d_{1\sigma}^{\dagger}d_{2\sigma}+d_{2\sigma}^{\dagger}d_{1\sigma}\right),\\
H_{Z} & = & \frac{\epsilon_{z}}{2}\sum_{l}(n_{l\uparrow}-n_{l\downarrow}),\\
H_{\mathrm{hf}} & = & \sum_{l}\mathbf{S}_{l}\cdot\mathbf{h}_{l};\,\,\,\,\,\mathbf{S}_{l}=\frac{1}{2}\sum_{\sigma\sigma^{\prime}}d_{l\sigma}^{\dagger}\pmb{\sigma}_{\sigma\sigma^{\prime}}d_{l\sigma^{\prime}}.\end{eqnarray}
 Here, $d_{l\sigma}^{\dagger}$ creates an electron with spin $\sigma$
in orbital state $l$ $(l=1,2)$, $V_{gl}$ is the single-particle
charging energy for orbital state $l$, $U$ is the two-particle charging
energy for two electrons in the same orbital state, and $U^{\prime}$
is the two-particle charging energy when there is one electron in
each orbital. When the orbital eigenstates are localized states in
quantum dot $l=1,2$, $V_{gl}$ is supplied by the back-gate voltage
on dot $l$ and $U$ $(U^{\prime})$ is the on-site (nearest-neighbor)
charging energy. $t_{12}$ is the hopping matrix element between the
two orbital states, $\epsilon_{z}$ is the electron Zeeman splitting,
$\mathbf{h}_{l}$ is the nuclear field (Overhauser operator) for an
electron in orbital $l$, and $\pmb{\sigma}_{\sigma\sigma^{\prime}}$
gives the matrix elements of the vector of Pauli matrices $\pmb{\sigma}=(\sigma_{x},\sigma_{y},\sigma_{z})$.
In the subspace of two electrons occupying two orbital states, the
spectrum of $H_{SP}+H_{C}$ consists of four degenerate {}``delocalized''
states with one electron in each orbital, all with unperturbed energy
$E_{(1,1)}$ (a singlet $\ket{S(1,1)}$ and three triplets: $\ket{T_{j}(1,1)};\, j=\pm,0$),
and two non-degenerate {}``localized'' singlet states $\ket{S(2,0)}$
and $\ket{S(0,2)}$, with two electrons in orbital $l=1$ or $l=2$,
having energy $E_{(2,0)}$ and $E_{(0,2)}$, respectively.

To derive an effective Hamiltonian $H_{\mathrm{eff}}$ from a given
Hamiltonian $H$, which has a set of nearly degenerate levels $\left\{ \ket{i}\right\} $,
we use the standard procedure\cite{stoneham:1975a},\begin{equation}
H_{\mathrm{eff}}=PHP+PHQ\frac{1}{E-QHQ}QHP,\label{eq:HeffPrescription}\end{equation}
 where $P=\sum_{i}\ket{i}\bra{i}$ is a projection operator onto the
relevant subspace and $Q=1-P$ is its complement. 

We choose the arbitrary zero of energy such that $E_{(1,1)}=V_{g1}+V_{g2}+U^{\prime}=0$
and introduce the detuning parameters \begin{eqnarray}
\delta_{1} & = & E_{(1,1)}-E_{(2,0)}=-2V_{g1}-U=-\delta-U-U^{\prime},\label{eq:delta1definition}\\
\delta_{2} & = & E_{(1,1)}-E_{(0,2)}=-2V_{g2}-U=\delta.\label{eq:delta2definition}\end{eqnarray}
 We then project onto the four-dimensional subspace formed by the
delocalized singlet $\ket{S(1,1)}$ and three delocalized triplet
states $\ket{T_{j}(1,1)},\,\,\,\,\, j=\pm,0$. That is, we choose
$Q=\ket{S(0,2)}\bra{S(0,2)}+\ket{S(2,0)}\bra{S(2,0)}$, $P=1-Q$.
When $\delta_{1},\delta_{2}\gg t_{12}$, we have \emph{$E\approx E_{(1,1)}=0$}
in the denominator of Eq. (\ref{eq:HeffPrescription}). This gives
an effective spin Hamiltonian in the subspace of one electron in each
orbital state:\begin{eqnarray}
H_{\mathrm{eff}} & = & \epsilon_{z}\sum_{l}S_{l}^{z}+\sum_{l}\mathbf{h}_{l}\cdot\mathbf{S}_{l}-J\left(\frac{1}{4}-\mathbf{S}_{1}\cdot\mathbf{S}_{2}\right),\\
 &  & J\approx-2t_{12}^{2}\left(\frac{1}{\delta}-\frac{1}{\delta+U+U^{\prime}}\right).\label{eq:ExchangeLargedelta}\end{eqnarray}
 This Hamiltonian is more conveniently rewritten in terms of the sum
and difference vectors of the electron spin and Overhauser operators
$\mathbf{S}=\mathbf{S}_{1}+\mathbf{S}_{2},\,\,\,\,\,\delta\mathbf{S}=\mathbf{S}_{1}-\mathbf{S}_{2}$
and $\mathbf{h}=\frac{1}{2}\left(\mathbf{h}_{1}+\mathbf{h}_{2}\right),\,\,\,\,\,\delta\mathbf{h}=\frac{1}{2}\left(\mathbf{h}_{1}-\mathbf{h}_{2}\right)$: 

\begin{eqnarray}
H_{\mathrm{eff}} & = & \epsilon_{z}S^{z}+\mathbf{h}\cdot\mathbf{S}+\delta\mathbf{h}\cdot\delta\mathbf{S}+\frac{J}{2}\mathbf{S}\cdot\mathbf{S}-J.\end{eqnarray}
 Neglecting the constant term, in the basis of singlet and three triplet
states, $\left\{ \ket{S(1,1)}=\ket{S},\ket{T_{j}(1,1)}=\ket{T_{j}},\, j=\pm,0\right\} $,
the Hamiltonian matrix for $H_{\mathrm{eff}}$ takes the form 

\begin{equation}
\left(\begin{array}{cccc}
0 & -\delta h^{+}/\sqrt{2} & \delta h^{z} & \delta h^{-}/\sqrt{2}\\
-\delta h^{-}/\sqrt{2} & J+\epsilon_{z}+h^{z} & h^{-}/\sqrt{2} & 0\\
\delta h^{z} & h^{+}/\sqrt{2} & J & h^{-}/\sqrt{2}\\
\delta h^{+}/\sqrt{2} & 0 & h^{+}/\sqrt{2} & J-\epsilon_{z}-h^{z}\end{array}\right),\end{equation}
 where $\delta h^{\pm}=\delta h^{x}\pm i\delta h^{y}$ and $h^{\pm}=h^{x}\pm ih^{y}$.
We are interested in this Hamiltonian in two limiting cases, where
it becomes block-diagonal in a two-dimensional subspace.

\subsection{Effective Hamiltonian in the $\ket{S}-\ket{T_{0}}$ subspace}

Projecting $H$ onto the two-dimensional subspace spanned by $\ket{T_{0}}\to\ket{\tau^{z}=+1}$
and $\ket{S}\to\ket{\tau^{z}=-1}$, we find\begin{equation}
H_{0}=N_{0}+\frac{1}{2}\mathbf{v}_{0}\cdot\pmb{\tau},\end{equation}
where $\pmb{\tau}=(\tau^{x},\tau^{y},\tau^{z})$ is a vector of Pauli
matrices. The leading and first subleading corrections to $H_{0}$
in powers of $1/\epsilon_{z}$ are ($H_{0}=H_{0}^{(0)}+H_{0}^{(1)}+\cdots$,
$H_{0}^{(i)}=N_{0}^{(i)}+\mathbf{v}_{0}^{(i)}$): \begin{eqnarray}
N_{0}^{(0)} & = & \frac{J}{2},\\
v_{0}^{z(0)} & = & J,\\
v_{0}^{+(0)} & = & 2\delta h^{z},\\
N_{0}^{(1)} & = & \frac{1}{4\epsilon_{z}}\left(\commute{h^{-}}{h^{+}}+\commute{\delta h^{-}}{\delta h^{+}}\right),\\
v_{0}^{z(1)} & = & \frac{1}{2\epsilon_{z}}\left(\commute{h^{-}}{h^{+}}-\commute{\delta h^{-}}{\delta h^{+}}\right),\\
v_{0}^{+(1)} & = & \frac{1}{\epsilon_{z}}\left(\delta h^{+}h^{-}+\delta h^{-}h^{+}\right).\end{eqnarray}
 Here, $\mathbf{N}_{X}=(N_{X}^{x},N_{X}^{y},N_{X}^{z})$, $\mathbf{v}_{X}=(v_{X}^{x},v_{X}^{y},v_{X}^{z})$,
$N_{X}^{\pm}=N_{X}^{x}\pm iN_{X}^{y}$, and $v_{X}^{\pm}=v_{X}^{x}\pm iv_{X}^{y}$.
For a typical unpolarized system, we estimate the size of all subleading
corrections from their r.m.s. expectation values, taken with respect
to an unpolarized nuclear state. This gives\begin{equation}
\left\langle H_{0}^{(1)}\right\rangle _{\mathrm{rms}}=O\left(\frac{\alpha^{2}}{\epsilon_{z}}\right),\end{equation}
 where $\alpha$ is given by $\alpha=\sqrt{I(I+1)}A/\sqrt{N}$ (for
a GaAs quantum dot containing $N\approx10^{5}$ nuclear spins, $1/\alpha\approx5\,\mathrm{ns}$).
We therefore expect dynamics calculated under $H_{0}^{(0)}$ to be
valid up to timescales on the order of $\epsilon_{z}/\alpha^{2}\gg1/\alpha$,
when $\epsilon_{z}\gg\alpha$.

\subsection{Effective Hamiltonian in the $\ket{S}-\ket{T_{+}}$ subspace}

When the Zeeman energy of the $\ket{T_{+}}$ triplet state approximately
compensates the exchange, $\max\left(\left\langle \mathbf{h}\right\rangle _{\mathrm{rms}},\left\langle \delta\mathbf{h}\right\rangle _{\mathrm{rms}},\left|\Delta\right|\right)\ll J$
(where $\Delta=\epsilon_{z}+J$), we find an effective Hamiltonian
in the subspace $\ket{T_{+}}\to\ket{\tau^{z}=+1},\,\ket{S}\to\ket{\tau^{z}=-1}$:\begin{equation}
H_{+}=N_{+}+\frac{1}{2}\mathbf{v}_{+}\cdot\pmb{\tau},\end{equation}
where the leading and subleading corrections in powers of $1/J$ are\begin{eqnarray}
N_{+}^{(0)} & = & \frac{1}{2}\left(\Delta+h^{z}\right),\\
v_{+}^{z(0)} & = & \Delta+h^{z},\\
v_{+}^{+(0)} & = & -\sqrt{2}\delta h^{+},\\
N_{+}^{(1)} & = & -\frac{1}{2J}\left(\left(\delta h^{z}\right)^{2}+\frac{1}{4}\delta h^{-}\delta h^{+}+\frac{1}{2}h^{-}h^{+}\right),\\
v_{+}^{z(1)} & = & \frac{1}{J}\left(\left(\delta h^{z}\right)^{2}+\frac{1}{4}\delta h^{-}\delta h^{+}-\frac{1}{2}h^{-}h^{+}\right),\\
v_{+}^{+(1)} & = & -\sqrt{2}\frac{\delta h^{z}h^{+}}{J}.\end{eqnarray}
 Once again, we estimate the influence of the subleading corrections
from their r.m.s. value with respect to a nuclear spin state of polarization
$p\ll1$, giving\begin{equation}
\left\langle H_{+}^{(1)}\right\rangle _{\mathrm{rms}}=O\left(\frac{\alpha^{2}}{J}\right).\end{equation}
 We therefore expect the dynamics under $H_{+}^{(0)}$ to be valid
up to time scales on the order of $t\sim J/\alpha^{2}\gg1/\alpha$
for $J\gg\alpha$.

\section{\label{sec:Asymptotics}Asymptotics}

\subsection{$C_{T_{0}}(\infty)$ for $J\gg2\sigma_{0},\, J\ll2\sigma_{0}$ }

In the limit of $J\to0$, we perform an asymptotic expansion of the
integral in Eq. (\ref{eq:CSaturationIntegral}) by separating the
prefactor into a constant piece and an unnormalized Lorentzian of
width $J/2$:\begin{equation}
C(x)=\frac{1}{2}\left(1-\frac{(J/2)^{2}}{\left(J/2\right)^{2}+x^{2}}\right).\end{equation}
 The Gaussian average over the constant term gives $1/2$ and when
$J/2\ll\sigma_{0}$, the typical $x$ contributing to the Lorentzian
part of Eq. (\ref{eq:CSaturationIntegral}) is $x\lesssim J/2\ll\sigma_{0}$,
so we approximate $\exp(-\frac{1}{2}x^{2}/\sigma_{0}^{2})\approx1$
in the integrand of this term. Integrating the Lorentzian then gives
the result in Eq. (\ref{eq:CorrelatorSaturationLimits}) for $J\ll2\sigma_{0}$.
In the opposite limit of $J\gg2\sigma_{0}$, the Lorentzian is slowly-varying
with respect to the Gaussian, and the prefactor can be expanded within
the integrand $C(x)\approx2x^{2}/J^{2}$. Performing the remaining
Gaussian integral gives the result in Eq. (\ref{eq:CorrelatorSaturationLimits})
for $J\gg2\sigma_{0}$.

\subsection{$C_{T_{0}}^{\mathrm{int}}(t)$ for $t\to\infty$}

To evaluate the integral in Eq. (\ref{eq:ComplexCT0}) at long times
when $J\ne0$, we make the change of variables $u=\sqrt{\lambda^{2}+\left(x/\sigma_{0}\right)^{2}}-\lambda$,
$\lambda=J/2\sigma_{0}$, $\tilde{t}=2\sigma_{0}t$, which gives\begin{widetext}

\begin{eqnarray}
\tilde{C}_{T_{0}}^{\mathrm{int}}(\tilde{t}/2\sigma_{0}) & =- & \frac{1}{\sqrt{2\pi}}\int_{0}^{\infty}du\frac{\sqrt{u(u+2\lambda)}}{u+\lambda}\exp\left\{ -\frac{1}{2}\left(u^{2}+2u\lambda\right)+i(u+\lambda)\tilde{t}\right\} ,\\
 &  & \lambda=J/2\sigma_{0},\,\,\,\,\,\tilde{t}=2\sigma_{0}t.\end{eqnarray}
 \end{widetext} At long times, the major contributions to this integral
come from a region near the lower limit, where $u\lesssim1/\tilde{t}$.
For $\tilde{t}\gg\max(1/\lambda,1)$ (i.e. $t\gg\max(1/J,1/2\sigma_{0})$),
we approximate the integrand by its form for $u\ll\max(\lambda,1)$,
retaining the exponential term as a cutoff. This gives

\begin{widetext}\begin{equation}
\tilde{C}_{T_{0}}^{\mathrm{int}}(\tilde{t}/2\sigma_{0})\sim-\frac{e^{i\lambda\tilde{t}}}{\sqrt{\pi\lambda}}\int_{0}^{\infty}du\sqrt{u}e^{-(\lambda-i\tilde{t})u}=-\frac{e^{i\lambda\tilde{t}}}{2\sqrt{\lambda}\left(\lambda-i\tilde{t}\right)^{3/2}}.\end{equation}

\end{widetext}When $\tilde{t}\gg\lambda$ (i.e. $t\gg J/4\sigma_{0}^{2}$),
we expand the denominator of the above expression, which gives the
result in Eq. (\ref{eq:CT0LongTimes}).

\bibliographystyle{apsrev}
\bibliography{/home/coish/papers/ddhfdecoherence/hfBibliography}

\end{document}